 \documentstyle[12pt]{article}

\let\a=\alpha \let\b=\beta   
    
\let\l=\lambda

 \let\W=\Omega
\let\la=\label  
  
\def\nn{\nonumber} \def\bd{\begin{document}} \def\ed{\end{document}}
\def\ds{\documentstyle} \let\fr=\frac \let\bl=\bigl \let\br=\bigr
\let\Br=\Bigr \let\Bl=\Bigl
\let\bm=\bibitem
\let\na=\nabla
\let\pa=\partial \let\ov=\overline
\newcommand{\be}{\begin{equation}}
\newcommand{\ee}{\end{equation}}
\def\ba{\begin{array}}
\def\ea{\end{array}}
\newcommand{\ho}[1]{$\, ^{#1}$}
\newcommand{\hoch}[1]{$\, ^{#1}$}
\newcommand{\bea}{\begin{eqnarray}}
\newcommand{\eea}{\end{eqnarray}}
\newcommand{\ra}{\rightarrow}
\newcommand{\lra}{\longrightarrow}
\newcommand{\Lra}{\Leftrightarrow}
\newcommand{\ap}{\alpha^\prime}
\newcommand{\bp}{\tilde \beta^\prime}
\newcommand{\tr}{{\rm tr} }
\newcommand{\Tr}{{\rm Tr} }
\newcommand{\NP}{Nucl. Phys. }
\newcommand{\tamphys}{\it Isaac Newton Institute for Mathematical Sciences\\
University of
Cambridge  \\ 20 Clarkson Road, Cambridge CB3 OEH, U.K.}

\newcommand{\auth}{M. J. Duff \footnote{On leave of absence from the Center for
Theoretical Physics,
 Texas A\&M University, College Station, Texas 77843. Research supported in
part by NSF
Grant PHY-9411543. }}

\begin{document}

\hfill{NI-94-015}

\hfill{CTP-TAMU-22/94}

\hfill{hep-th/9410046}

\vspace{24pt}

\begin{center}
{ \large {\bf KALUZA-KLEIN THEORY IN PERSPECTIVE \footnote{Talk delivered at
the Oskar Klein
 Centenary  Nobel Symposium, Stockholm, September 19-21, 1994.  }}}

\vspace{36pt}

\auth

\vspace{10pt}

{\tamphys}

\vspace{44pt}

\underline{ABSTRACT}

\end{center}

The Kaluza-Klein idea of extra spacetime dimensions continues to pervade
current attempts to unify
the fundamental forces, but in ways somewhat different from that originally
envisaged. We present
a modern perspective on the role of internal dimensions in physics, focussing
in particular on
superstring theory. A novel result is the interpretation of Kaluza-Klein string
states as extreme
black holes.

{\vfill\leftline{}\vfill
\leftline{October 1994}
\vskip 10pt
\baselineskip=24pt
\pagebreak

\section{The Kaluza idea}

Cast your minds back to 1919.  Maxwell's theory of electromagnetism was well
established and Einstein had recently formulated his General Theory of
Relativity.
By contrast, the strong and weak interactions were not well understood.  In
searching for a unified
theory of the fundamental forces, therefore, it was natural to attempt to merge
gravity with
electromagnetism. This Kaluza \cite{Kaluza} was able to do through the
remarkable device of
postulating an extra fifth dimension for spacetime. Consider Einstein's theory
of pure gravity in
five spacetime dimensions with signature $(-,+,+,+,+)$. The line element is
given by
\begin{equation}
d\hat{s}^2=\hat{g}_{\hat{\mu}\hat{\nu}}d\hat{x}^{\hat{\mu}}d\hat{x}^{\hat{\nu}}
\label{Einstein}
\end{equation}
where $\hat{\mu} = 0,1,2,3,4$ and all hatted quantities are five-dimensional.
Kaluza then made
the $4+1$ split
\begin{equation}
\hat{g}_{\hat{\mu}\hat{\nu}}=e^{\phi/\sqrt{3}}
\pmatrix{g_{\mu\nu}+e^{-\sqrt{3}\phi}A_{\mu}A_{\nu}&e^{-\sqrt{3}\phi}A_{\mu}\cr
e^{-\sqrt{3}\phi}A_{\nu}&e^{-\sqrt{3}\phi}}
\label{metric}
\end{equation}
where $\hat{x}^{\hat{\mu}}=(x^{\mu},y)$, $\mu = 0,1,2,3,$ and all unhatted
quantities are
four-dimensional.  Thus the fields $g_{\mu\nu}(x)$, $A_{\mu}(x)$ and $\phi (x)$
transform
respectively as a tensor, a vector and a scalar under four-dimensional general
coordinate
transformations.  All this was at the classical level, of course, but in the
modern parlance of
quantum field theory, they would be described as the spin $2$ graviton, the
spin $1$ photon and
the spin $0$ dilaton\footnote{This was considered an embarassment in 1919, and
was
(inconsistently) set equal to zero. However, it was later revived by Jordan
\cite{Jordan} and
Thiry \cite{Thiry} and subsequently stimulated Brans-Dicke \cite {Brans}
theories of gravity.
As we shall see, the dilaton also plays a crucial role in superstring theory}.
Of course it is
not enough to call $A_{\mu}$ by the name photon, one must demonstrate that it
satisfies Maxwell's
equations and here we see the Kaluza miracle at work.   After making the same
$4+1$ split of the
five-dimensional Einstein equations $\hat{R}_{\hat{\mu}\hat{\nu}}=0$, we
correctly recover the
not only the Einstein equations for $g_{\mu\nu}(x)$ but also the Maxwell
equation for
$A_{\mu}(x)$ and the massless Klein-Gordon equation for $\phi (x)$.  Thus
Maxwell's theory of electromagnetism is an inevitable consequence of Einstein's
general theory of
relativity, given that one is willing to buy the idea of a fifth dimension.

\section{The Klein idea}

Attractive though Kaluza's idea was, it suffered from two obvious drawbacks.
First, although the
indices were allowed to range over $0,1,2,3,4$, for no very good reason the
dependence on the extra
coordinate $y$ was suppressed. Secondly, if there is a fifth dimension why
haven't we seen it? The
resolution of both these problems was supplied by Oskar Klein \cite{Klein1} in
1926.
Klein insisted on treating the extra dimension seriously but assumed the fifth
dimension to have
circular topology so that the coordinate $y$ is periodic, $0\leq my \leq 2\pi$,
where $m$ is
the inverse radius of the circle $S^1$. Thus the space has topology $R^4 \times
S^1$.
It is difficult to envisage a spacetime with this topology but a simpler
analogy is
provided by a hosepipe: at large distances it looks like a line $R^1$ but
closer inspection
reveals that at every point on the line there is a little circle, and the
topology is $R^1 \times
S^1$.  So it was that Klein suggested that there is a little circle at each
point in four-dimensional
spacetime.

Let us consider Klein's proposal from a modern perspective. We start with pure
gravity in five
dimensions described by the action
\begin{equation}
\hat S=\frac{1}{2\hat  \kappa{}^2}\int d^5\hat{x}\sqrt{-\hat{g}}\hat{R}
\label{action}
\end{equation}
$\hat S$ is invariant under the
five-dimensional general coordinate transformations
\begin{equation}
\delta \hat g_{\hat \mu\hat \nu}= \partial_{\hat \mu} \hat \xi^{\hat \rho}\hat
g_{\hat \rho\hat \nu} +\partial_{\hat \nu} \hat \xi^{\hat \rho}\hat g_{\hat
\rho\hat \mu}
+\hat \xi^{\hat \rho}\partial_{\hat \rho}\hat g_{\hat \mu\hat \nu}
\label{general}
\end{equation}
The periodicity in $y$ means that the fields $g_{\mu\nu}(x,y)$, $A_{\mu}(x,y)$
and $\phi
(x,y)$ may be expanded in the form
\[
 g_{\mu\nu}(x,y)= \sum_{n=-\infty}^{n=\infty} g_{\mu\nu n}(x)e^{inmy},
\]
\[
A_{\mu}(x,y)=\sum_{n=-\infty}^{n=\infty}A_{\mu
n}(x)e^{inmy},
\]
\be
\phi(x,y)=\sum_{n=-\infty}^{n=\infty}\phi_ne^{inmy}
\label{fourier}
\ee
with
\begin{equation}
 g^*{}_{\mu\nu n}(x)=g_{\mu\nu -n}(x)
\end{equation}
etc. So (as one now finds in all the textbooks) a Kaluza-Klein theory describes
an infinite
number of four-dimensional {\it fields}. However (as one finds in none of the
textbooks) it also
describes an infinite number of four-dimensional {\it symmetries} since we may
also Fourier expand
the general coordinate parameter $\hat \xi^{\hat \mu}(x,y)$ as follows
\[ \hat
\xi^{\mu}(x,y)=\sum_{n=-\infty}^{n=\infty} \xi^{\mu}{}_n(x)e^{inmy} \]
\begin{equation}
\hat  \xi^{4}(x,y)=\sum_{n=-\infty}^{n=\infty} \xi^{4}{}_n(x)e^{inmy}
\label{fourier2}
\end{equation}
with $\hat \xi^*{}^{\hat\mu}{}_n=\hat \xi^{\hat\mu}{}_{-n}$.

Let us first focus on the $n=0$ modes in (\ref{fourier}) which are  just
Kaluza's
graviton, photon and dilaton. Substituting (\ref{metric}) and  (\ref{fourier})
in the action
(\ref{action}), integrating over $y$ and retaining just the $n=0$ terms we
obtain (dropping the
$0$ subscripts) \begin{equation}
 S=\frac{1}{2 \kappa{}^2}\int
d^4{x}\sqrt{{-g}}[R-\frac{1}{2}\partial_{\mu}\phi\partial^{\mu}\phi
-\frac{1}{4}e^{-\sqrt{3}\phi}F_{\mu\nu}F^{\mu\nu}]
\label{action2}
\end{equation}
where $2\pi\kappa^2=m\hat \kappa^2$ and
$F_{\mu\nu}=\partial_{\mu}A_{\nu}-\partial_{\nu}A_{\mu}$.
{}From (\ref{general}), this action is invariant under general coordinate
transformations with
parameter $\xi^{\mu}{}_0$, i.e (again dropping the $0$ subscripts) \[
\delta  g_{ \mu \nu}= \partial_{ \mu}  \xi^{ \rho}
g_{ \rho \nu} +\partial_{ \nu}  \xi^{ \rho}g_{ \mu\rho}
+\xi^{\rho}\partial_{\rho}g_{\mu\nu}
\]
\[
\delta A_{\mu}=\partial_{\mu} \xi^{\rho}A_{\rho}+\xi^{\rho}\partial_{\rho}
A_{\mu}
\]
\begin{equation}
\delta \phi=\xi^{\rho}\partial_{\rho}\phi,
\label{general2}
\end{equation}
local gauge transformations with parameter $\xi^{4}{}_0$
\begin{equation}
\delta A_{\mu}=\partial_{\mu}\xi^{4}
\label{gauge}
\end{equation}
and global scale transformations with parameter $\lambda$
\begin{equation}
\delta A_{\mu}=\lambda A_{\mu},\,\,\,
\delta \phi=-2\lambda/\sqrt 3
\label{scale}
\end{equation}
The symmetry of the vacuum, determined by the VEVs
\begin{equation}
<g_{ \mu \nu}>=\eta_{ \mu \nu},\,\,\,<A_{\mu}>=0,\,\,\,<\phi>=\phi_0
\label{vevs}
\end{equation}
is the four-dimensional Poincare group $\times
R$. Thus, the masslessness of the the graviton is due to general covariance,
the masslessness of
the photon to gauge invariance, but the dilaton is massless because it is the
Goldstone boson
associated with the spontaneous breakdown of the global scale invariance. Note
that the gauge
group is $R$ rather than $U(1)$ because this truncated $n=0$ theory has lost
all memory of the
periodicity in $y$.

Now, however, let us include the $n\neq0$ modes. An important observation is
that the assumed
topology of the ground state, namely $R^4 \times S^1$ restricts us to general
coordinate
transformations periodic in $y$.  Whereas the general covariance
(\ref{general2}) and local
gauge invariance (\ref{gauge}) simply correspond to the $n=0$ modes of
(\ref{general})
respectively, the global scale invariance is no longer a symmetry because it
corresponds to a
rescaling
\begin{equation}
\delta \hat g_{\hat \mu \hat \nu}= -\frac{1}{2}\lambda \hat g_{\hat \mu \hat
\nu}
\end{equation}
combined with a a general coordinate transformation
\begin{equation}
\xi^4=-\lambda y/m
\end{equation}
which is now forbidden by the periodicity requirement.  The field $\phi_0$ is
therefore merely a
pseudo-Goldstone boson.

Just as ordinary general covariance may be regarded as the local gauge symmetry
corresponding to
the global Poincare algebra  and local gauge invariance as the gauge symmetry
corresponding to
the  global abelian algebra, so the infinite parameter local transformations
(\ref{fourier2})
correspond to an infinite-parameter global algebra with generators
\[
P^{\mu}{}_n=e^{inmy}\partial^{\mu}
\]
\[
M^{\mu\nu}{}_n=e^{inmy}(x^{\mu}\partial^{\nu}-x^{\nu}\partial^{\mu})
\]
\be
Q_n=ie^{inmy}\partial/\partial(my)
\ee
It is in fact a Kac-Moody-Virasoro generalization of the Poincare/gauge algebra
\cite{Dolan}.
Although this larger algebra describes a symmetry of the four-dimensional
theory, the symmetry
of the vacuum determined by (\ref{vevs}) is only Poincare $\times U(1)$.  Thus
the gauge
parameters $\xi^{\mu}{}_n$ and $\xi^{4}{}_n$ with $n\neq 0$ each correspond to
spontaneously
broken generators, and it follows that for $n\neq 0$ the fields $A_{\mu}{}_n$
and $\phi_n$ are the
corresponding Goldstone boson fields. The gauge fields $g_{\mu\nu}{}_n$, with
two degrees of
freedom, will then each acquire a mass by absorbing the the two degrees of
freedom of each vector
Goldstone boson $A_{\mu}{}_n$ and the one degree of freedom of each scalar
Goldstone boson
$\phi_n$ to yield a pure spin $2$ massive particle with five degrees of
freedom.  This accords
with the observation that the massive spectrum is pure spin two \cite{Salam1}.
Thus we
find an infinite tower of charged, massive spin $2$ particles with charges
$e_n$ and masses $m_n$
given by
\begin{equation}
e_n=n\sqrt 2\kappa m,\,\,\, m_n=|n|m
\label{charge}
\end{equation}
Thus Klein explained (for the first time) the quantization of electric
charge \cite{Klein2}. (Note also that charge conjugation is just parity
tranformation $y \rightarrow
-y$ in the fifth dimesion.)  Of course, if we identify the fundamental unit of
charge $e=\sqrt
2\kappa m$ with the charge on the electron, then we are forced to take $m$ to
be very large: the
Planck mass $10^{19}$ $GeV$, way beyond the range of any current or forseeable
accelerator. This
answers the second question left unanswered by Kaluza because with $m$ very
large, the radius of the
circle must be very small: the Planck size $10^{-35}$ $meters$, which
satisfactorily accords with our
everyday experience of living in four spacetime dimensions.

It is interesting to note that, despite the inconsistency problems
\cite{Boulware} that arise in
coupling a finite number of massive spin two particles to gravity and/or
electromagnetism,
Kaluza-Klein theory is consistent by virtue of having an {\it infinite} tower
of such states.  Any
attempt to truncate to a finite non-zero number of massive modes would
reintroduce the inconsistency
\cite{Stelle}. We also note, however, that these massive  Kaluza-Klein modes
have the unusual
gyromagnetic ratio $g=1$ \cite{Hosoya}, which seems to lead to unacceptable
high-energy behaviour
for Compton scattering \cite{Ferrara}.  Moreover, as we shall see in section
(\ref{strings}), where
we embed the theory in a superstring theory, these Kaluza-Klein states will
persist as a subset of
the full string spectrum.  However, string theory comes to the rescue and
ensures correct
high-energy behaviour.

In summary, it seems that a five-dimensional world with one of its dimensions
compactified on a
circle is operationally indistinguishable from a four dimensional world with a
very particular
(albeit infinite) mass spectrum. From this perspective, therefore, it seems
that one could kick the
ladder away and forget about the fifth dimension.

\section{The Kaluza-Klein black hole}
\la{black}
The equations which follow from (\ref{action2}) admit electrically charged
black hole solutions \cite{Dobiasch,Chodos,Pollard,Gibbons}:
\[
ds^2=-\Delta_+\Delta_-{}^{-1/2}dt^2+\Delta_+{}^{-1}\Delta_-{}^{1/2}dr^2+r^2\Delta_-{}^{3/2}
d\Omega^2 \]
\[
e^{2\phi}=\Delta_-^{\sqrt 3}
\]
\be
e^{-\sqrt 3 \phi}{}^*F=(r_+r_-)^{1/2}\epsilon_2
\la{solution}
\ee
where $\Delta_{\pm}=1-r_{\pm}/r$ and $ \epsilon_2$ is the volume form on $S^2$.
 The electric
charge $e$ and ADM mass $m$ are related to $r_{\pm}$ by
\[
{\sqrt 2 \kappa}e/4\pi=(r_+r_-)^{1/2}
\]
\be
{2 \kappa^2}m/4\pi=2r_+ - r_-
\ee
The existence of an event horizon, $r_+\geq r_-$, thus implies the bound
\be
\sqrt{2} \kappa m \geq e
\la{bound}
\ee
In the extreme limit, $r_+=r_-$, the line element reduces
to  \be
ds^2=-\Delta_-{}^{1/2}dt^2+\Delta_-{}^{-1/2}dr^2+r^2\Delta_-{}^{3/2} d\Omega^2
\la{extreme}
\ee
and the bound (\ref{bound}) is saturated. Note that this yields exactly the
same charge
to mass ratio (\ref{charge}) as the massive Kaluza-Klein states.  As we shall
show in section
(\ref{strings}), this is no coincidence: the massive states {\it are} extreme
black holes!

The same equations also admit the magnetically charged black hole solution with
the same metric but
with \[
e^{-2\phi}=\Delta_-^{\sqrt 3}
\]
\be
F=(r_+r_-)^{1/2}\epsilon_2
\ee
 and with magnetic charge $g$ given by
\be
{\sqrt 2 \kappa}g/4\pi=(r_+r_-)^{1/2}
\ee
In the extreme limit, $r_+=r_-$, this is the Kaluza-Klein monopole
\cite{Pollard,Perry,Sorkin}.

We note that the four-dimensional monopole metric $g_{\mu\nu}$ of
(\ref{extreme}) exhibits a
curvature singularity at $r=r_-$, even though the five-dimensional metric $\hat
g_{\hat \mu \hat
\nu}$ of (\ref{metric}) from which it is descended is perfectly regular! This
appears to contradict the impression gained at the end of the last section that
the
five-dimensional perspective is an unnecessary luxury.  However, consider the
Weyl rescaled metric
$\tilde g_{\mu\nu}=e^{\sqrt 3 \phi}g_{\mu\nu}$. The magnetic monopole line
element is now
\be
d\tilde s^2=-\Delta_-{}^{-1}dt^2+\Delta_-{}^{-2}dr^2+r^2 d\Omega^2
\ee
and the curvature singularity at $r=r_-$ has disappeared! The physical
significance of this
metric is that it is the one that couples to the worldline of an {\it
electrically} charged point
particle \cite{Lu1,Khuri}.  We shall return to this in section (\ref{strings}).

\section{The humble torus}
\la{torus}
In $D>5$ dimensions the pure gravity field equations $\hat R_{\hat \mu \hat
\nu}=0$ are
consistent with the ground state $M_4 \times T^k$ where $T^k$ is the metrically
flat $k$-torus
$T^k= S^1 \times S^1 \times ...S^1$.  The gauge group is now $G=[U(1)]^k$.  The
count of
massless modes (degrees of freedom) is: $1$ spin $2$ $(2)$; $k$ spin $1$
$(2k)$; $k(k+1)/2$
spin $0$ $(k(k+1)/2)$.  Note
that the total number of degrees of freedom is $(4+k)(1+k)/2$ which matches the
degrees of
freedom of a graviton in $D=4+k$ dimensions. The number of scalars is
given by the moduli of $T^k$ and they parameterize the non-linear
$\sigma$-model
$GL(k,R)/SO(k)$ \cite{Cremmer}. Thus even the {\it humble torus} (in the words
of Abdus Salam), the
simplest of extra-dimensional geometries one could envisage, gives rise to a
non-trival
four-dimensional world.

As we shall now show, the torus becomes even more
non-trivial in the context of supergravity theory. Here, in addition to the
metric $\hat g_{\hat
\mu \hat \nu}$, the $D=10$ supergravity multiplet contains a 2-form $\hat
B_{\hat \mu\hat \nu}$
and a dilaton $\hat \Phi$.  After compactification to $D=4$ on a torus, the
bosonic degrees
of freedom count now is: $1$ spin $2$; $12$ spin $1$; and $38$ scalars composed
of $36$ moduli
parameterizing $SO(6,6)/{SO(6) \times SO(6)}$ and an axion and dilaton
parameterizing
$SL(2,R)/U(1)$. There will also be an equal number of fermion degrees of
freedom: $4$ spin $3/2$
and $28$ spin $1/2$. If we include the Yang-Mills multiplet we obtain a further
$16$ spin $1$;
$64$ spin $1/2$ and $96$ spin $0$ so that the moduli coset is then
$SO(6,22)/{SO(6) \times
SO(22)}$. Since this theory is the field theory limit of the heterotic string
compactfied on a
generic torus  let us consider the action in more detail \cite{Schwarz,Sen}.
Its bosonic sector is
given by:    %
\bea
\lefteqn{ S=\frac{1}{2\kappa^2}\int d^4x\sqrt{-G}e^{-\Phi}[R_G +
G^{\mu\nu}\partial_{\mu}\Phi\partial_{\nu}\Phi
-\frac{1}{12}G^{\mu\lambda}
 G^{\nu\tau}G^{\rho\sigma}H_{\mu\nu\rho}H_{\lambda\tau\sigma}}\nn\\&&
-\frac{1}{4}G^{\mu\lambda}G^{\nu\tau}F_{\mu\nu}{}^a(LML)_{ab}
   F_{\lambda\tau}{}^b+ \frac{1}{8}
     G^{\mu\nu}Tr(\partial_{\mu}ML\partial_{\nu}ML)]
\la{3}
\eea
where $F_{\mu\nu}{}^a=\partial_{\mu}A_{\nu}{}^a-\partial_{\nu}A_{\mu}{}^a$
and
$H_{\mu\nu\rho}=(\partial_{\mu}B_{\nu\rho}+2A_{\mu}{}^aL_{ab}F_{\nu\rho}{}^b)
+ {\rm permutations}$.  Here $\Phi$ is the $D=4$ dilaton, $R_G$ is the
scalar curvature formed from the string metric $G_{\mu\nu}$, related to the
canonical metric $g_{\mu\nu}$ by $G_{\mu\nu}\equiv e^{\Phi}g_{\mu\nu}$.
$B_{\mu\nu}$ is the 2-form which couples to the string worldsheet and
$A_{\mu}{}^a$ ($a=1,...,28$) are the abelian gauge fields. $M$ is a
symmetric $28\times28$ dimensional matrix of scalar fields satisfying
$MLM=L$ where $L$ is the invariant metric on $O(6,22)$:
\be
L=\pmatrix{0&I_6&0\cr I_6&0&0 \cr 0&0&-I_{16}}.
\la{L}
\ee
The action is invariant under the $O(6,22)$ transformations
$M\rightarrow\Omega M\Omega^T$,
$A_{\mu}{}^a\rightarrow\Omega^{a}{}_{b}A_{\mu}{}^b$, $G_{\mu\nu}\rightarrow
G_{\mu\nu}$, $B_{\mu\nu}\rightarrow B_{\mu\nu}$, $\Phi\rightarrow\Phi$,
where $\Omega$ is an $O(6,22)$ matrix satisfying $\Omega^TL\Omega=L$.
{\it $T$-duality} corresponds to the $O(6,22;Z)$ subgroup and is known to be an
exact symmetry of the full string theory. The equations of motion, though
not the action, are also invariant under the $SL(2,R)$ transformations:
${\cal M}\rightarrow \omega{\cal M}\omega^T,{\cal F}_{\mu\nu}{}^{a\alpha}
  \rightarrow  \omega^{\alpha}{}_{\beta}{\cal F}_{\mu\nu}{}^{a\beta},
 g_{\mu\nu}\rightarrow g_{\mu\nu},\,\,M\rightarrow M$
where $\a=1,2$ with ${\cal F}_{\mu\nu}{}^{a1}=F_{\mu\nu}{}^{a}$ and ${\cal
F}_{\mu\nu}{}^{a2}=\left(\lambda_2(ML)^a{}_{b}\tilde F_{\mu\nu}{}^{b}+
\lambda_1 F_{\mu\nu}{}^{a}\right)$, where $\omega$ is an $SL(2,R)$ matrix
satisfying  $\omega^T{\cal L}\omega={\cal L}$ and where
\be
{\cal M}=\frac{1}{\lambda_2}\left(\begin{array}{cc}
1&\lambda_1\\
\lambda_1&|\lambda|^2
\end{array}\right),\,\,\,
{\cal L}=\left(\begin{array}{cc}
0&1\\
-1&0\end{array}\right).
\la{4}
\ee
$\l$ is given by $\lambda=\Psi+ie^{-\Phi}\equiv\lambda_1+i\lambda_2$. The
axion $\Psi$ is defined through the relation
$\sqrt{-g}H^{\mu\nu\rho}=-e^{2\Phi}
\epsilon^{\mu\nu\rho\sigma}\partial_{\sigma}\Psi$.  {\it $S$-duality}
corresponds
to the $SL(2,Z)$ subgroup and, as discussed in section (\ref{strings}) there is
now a good deal of
evidence
\cite{Duff1,Duff2,Font,Kalara,Schwarz,Sen,Binetruy,Khuri2,Rahmfeld,Gauntlett,Vafa} in favor
of its also being an exact symmetry of the full string theory. It generalizes
earlier conjectures in
global Yang-Mills theories \cite{Montonen,Goddard,Witten,Osborn}

\section{Non-abelian generalization}
\la{nonabelian}
The arrival of Yang-Mills gauge theories in 1954 presented an altogether
different challenge to
higher-dimensional gravity theories: could they also account for {\it
non-abelian} gauge bosons?
Curiously enough, Oskar Klein \cite{Klein3} came close\footnote{See the article
by Gross
\cite{Gross}.} to discovering non-abelian gauge fields in 1939 while
investigating $D=5$
geometries, but their significance was never fully articulated.  The first
concrete attempt seems
to be that of De Witt \cite{DeWitt} in 1963 and this was followed by work of
Rayski \cite{Rayski},
Kerner \cite{Kerner}, Trautmann \cite{Trautmann}, Cho \cite{Cho}, Cho and
Freund \cite{Freund},
Cho and Jang \cite{Jang} and others. The breakthrough was the realization
that the gauge group  $G$ obtained in $D=4$ was connected to the {\it isometry
group} of the extra
dimensions which, in analogy with $S^1$, were taken to be compact to ensure the
compactness of $G$.
Thus, it was argued, $SU(2)$ gauge bosons arose from taking three extra
dimensions  and assigning
to them the geometry of a three-sphere which was, after all, the $SU(2)$ group
manifold.

With the wisdom of hindsight, we can now identify several shortcomings of these
non-abelian
developments. First, little attention was paid to the question of why the extra
dimensions were
compactified and whether this was consistent with the higher-dimensional field
equations.  It
was usually a completely ad hoc procedure.  This was remedied by the idea of
{\it spontaneous
compactification}.  Here one looks for stable {\it ground state} solutions of
the field
equations for which the metric describes a product manifold $M_4 \times M_k$
where $M_4$ is
four-dimensional spacetime with the usual signature and $M_k$ is a compact
"internal" space
with Euclidean signaature. As shown by Cremmer {\it et al} \cite{Cremmer2} it
was necessary to
augment pure gravity with matter fields in order to achieve a satisfactory
compactification.   A
second shortcoming was the failure to realize that the the extra-dimensional
manifold need not
correspond to a group space $G$ in order to obtain Yang-Mills gauge fields with
$G$ as their gauge
group.  Now we know that any $M_k$ with $G$ as its isometry group will do,
i.e., any metric
admitting the Killing vectors of $G$.  This could be a homogeneous space. In
this case the group $G$
acts transitively and we may write the manifold as the coset space $M_k=G/H$
where $H$ is the
isotropy subgroup of the isometry group $G$.  The use of such homogeneous
spaces in Kaluza-Klein
theories was was discussed by Luciani \cite{Luciani}. Since $k= dim G-dim H$,
one was no longer
obliged to have only one gauge boson for each extra dimension, as had
previously been assumed.
Indeed, the isometry group of a group manifold can be as large as $G \times G$
if we use the
bi-invariant metric, so $S^3$ can give $SU(2) \times SU(2)$ gauge bosons and
not merely SU(2).

Deriving Yang-Mills fields from gravity is perhaps the most beautiful aspect of
Kaluza-Klein
theories
so let us examine how it works. Let us make the $4+k$ split $\hat x ^{\hat
\mu}=(x^{\mu},y^n)$, $
\hat \mu=0,1,...,D-1$,  $\mu=0,1,2,3$ and $n=4,5,...,D-1$.  Consider the
$D$-dimensional metric $\hat
g_{\hat \mu \hat \nu}(x,y)$ and in particular, the off diagonal component $\hat
g_{ \mu n}(x,y)$.
In its Fourier expansion, the lowest term looks like
\be
\hat g_{ \mu n}(x,y)=A_{\mu}{}^i(x)K_n{}^i(y)+...
\la{killing}
\ee
where $K^i=K^{ni}\partial_n$ is a Killing vector obeying the Lie algebra of $G$
\be
[K^i,K^j]=f^{ij}{}_{k}K^k
\ee
and $f^i{}_{jk}$ are the structure constants.  Now we
consider the general coordinate transformation (\ref{general}) and focus our
attention on the
very special transformation \be
\hat \xi^{\hat \mu}(x,y)= (0,\epsilon^i(x)K^{mi}(y))
\ee
with $\epsilon^i(x)$ arbitrary. Then from (\ref{general}) we may compute the
transformation rule
for $\hat g_{\mu n}(x,y)$ and hence from (\ref{killing}) that for
$A_{\mu}{}^i(x)$.  We find
\be
\delta  A_{\mu}{}^i(x)= \partial_{\mu} \epsilon^i(x)
-f^i{}_{jk}A_{\mu}{}^j(x)\epsilon^k(x)
\ee
This is precisely the transformation law for a Yang-Mills field with gauge
group $G$.  Hence $G$
is a subgroup of the $D$-dimensional general coordinate group.  In summary, the
basic idea is that
what we perceive to be {\it internal symmetries} in four dimensions are really
{\it spacetime
symmetries} in the extra dimensions. Carrying this logic to its ultimate
conclusion, one might be
tempted to conclude that there is no such thing in nature as an internal
symmetry, even apparent
discrete internal symmetries like charge conjugation being just discrete
spacetime tranformations
in the extra dimensions.

One can only speculate on how the course of twentieth century physics might
have changed if, in
groping towards non-abelian gauge fields in 1939 \cite{Gross}, Klein had
applied his own ideas to a
{\it sphere} instead of a circle.

\section{Kaluza-Klein Supergravity}
\la{supergravity}

The history of Kaluza-Klein took a totally new turn with the advent of
supergravity
\cite{Freedman,Deser} whose mathematical consistency requires $D\leq 11$
dimensions \cite{Nahm}.
Cremmer, Julia and Scherk \cite{Scherk} were able to construct the $N=1,D=11$
supergravity
Lagrangian, describing the interaction of the elfbein $e_M{}^A$, the gravitino
$\Psi_M$, and
the three-index gauge field $A_{MNP}$. The uniqueness of the field equations
meant that not only
did Kaluza-Klein enthusiasts have a guide to the dimensionality of spacetime
but also a guide to
the correct interactions with gravity and matter. In $D\leq 11$, supersymmetric
theories are no
longer unique but still very restrictive. It seemed that this restriction on
the
dimension and the restrictions on the interactions, which supersymmetry
provides, were essential
to any successful Kaluza-Klein unification.  Otherwise one is wandering in the
wilderness: the
problem of finding the Lagrangian of the world in $D=4$ is simply replaced by
the problem of
finding the the Lagrangian of the world in $D>4$ with the extra headache of
which $D$ to pick.
Nothing has been gained by way of economy of thought.  Weinberg \cite{Weinberg}
is fond of
recalling the fable of the ''stone soup" when discussing this problem.  Just as
the promise of
delicious soup made from stones proved to mean stones plus meat and vegetables,
so the
Kaluza-Klein promise of a unified theory made only from gravity had proved to
mean gravity plus
a whole variety of matter fields with each author choosing his favorite
ingredients.  But by
combining gravity and matter into one simple superfield to which no further
supermatter may be
added, $D=11$ realized Einstein's old dream of replacing the ''base wood" of
matter by the ''pure
marble" of geometry.  Eleven-dimensional supergravity is ''marble soup".

The early $1980$s thus
marked a major renaissance of extra dimensions. First Freund and Rubin
\cite{Rubin} showed
that the $3$-form of $D=11$ supergravity provided a dynamical mechanism whereby
$7$ of the $11$
dimensions compactify spontaneously. Then Witten \cite{Witten2} considered the
search for a
realistic Kaluza-Klein theory pointing out both the advantages of $D=11$
supergravity ( seven
extra dimensions is the minimum to accommodate $SU(3) \times SU(2) \times
U(1)$) and its
disadvantages ( quarks and leptons do not fit into the right representations,
in particular the
four dimensional theory cannot be {\it chiral}). Thirdly, Salam and Strathdee
\cite{Salam1} laid
the foundations for much of the subsequent research on harmonic expansions on
coset spaces,
essential for the understanding of the massive modes.  All this sparked off the
observation
\cite{Pope} that the seven extra dimensions of $D=11$ supergravity could, via
the Freund-Rubin
ansatz yield ground-state solutions of the form $(D=4$ Anti de Sitter space)
$\times S^7$ and
that since $S^7$ has isometry group $SO(8)$, this would give rise to a $D=4$
theory with $SO(8)$
invariance.  It was not difficult to prove that $S^7$ also admits $8$ {\it
Killing spinors} and
hence gives rise to $N=8$ supersymmetry in $D=4$. It was therefore natural to
conjecture
\cite{Pope,Duff3,Nilsson1} that the massless supermultiplet of spins
$(2,3/2,1,1/2,0^+,0^-)$ in the
$SO(8)$ representations $(1,8,28,56,35,35)$ corresponded to the gauged $N=8$
theory of de Wit and
Nicolai \cite{deWit}.

There then followed a deluge of activity in both $D=11$ and $D<11$ Kaluza-Klein
supergravity (and in
Kaluza-Klein cosmology and quantum effects in Kaluza-Klein theories) by
Abbott, Alvarez, Appelquist, Aurelia, Awada, Bais, Barkanroth, Barr, Bars,
Bele'n Gavela, Berkov,
Biran, Candelas, Castellani, Ceresole, Chapline, Cho, Chodos, Coquereax,
D'Auria, De Alwis, Dereli,
Detweiler, deWit, Duff, Ellis, Emel'yanov, Englert, Fre, Freedman, Freund,
Fujii, Gell-Mann, Giani,
Gibbons, Gunaydin, Gursey, Hurni, Jadcyk, Inami, Ito, Kato, Kogan, Koh, Kolb,
Lukierski, Maeda,
Manton, MacDowell, McKenzie, Mecklenburg, Minnaert, Moorhouse, Morel, Nicolai,
Nikitin, Nilsson,
Nixon, Ohta, Orzalesi, Page, Panahhimoghaddam, Pilch, Pollard, Pope, Randjbar
Daemi, Romans, Rooman,
Rozenthal, Rubakov, Salam, Sahdev, Schaposnikov, Sezgin, Shafi, Slansky,
Sorokin, Spindel,
Strathdee, Sudbery, Tanii, Tkach, Toms, Townsend, Tucker, Tze, van Baal, van
Nieuwenhuizen, Volkov,
Voronov, Wang, N. Warner, R. Warner, Weinberg, West, Wetterich, Wu, Yamagishi,
Yasuda, Zwiebach, to
name but some.

\section{Superstrings}
\la{1984}

Thus up until the summer of 1984 various proposals were put forward combining
supersymmetry and the
Kaluza-Klein idea but none with complete success. Those based on conventional
field theory
suffered from various problems, not least of which was the traditional
objection to a
non-renormalizable theory of gravity.  Those based on superstrings seemed
better from this point of
view, and also from the point of view of chirality, but had problems of their
own. The
realistic-looking strings appeared to suffer from inconsistencies (anomalies
akin
to the triangle anomalies of the standard model) while the anomaly-free strings
did not appear
realistic. In particular, they seemed to live in ten spacetime dimensions
rather than undergoing a
spontaneous compactification to four spacetime dimensions as demanded by the
Kaluza-Klein idea.
This was the sorry state of affairs until the September 1984 superstring
revolution:

1) Green and Schwarz \cite{Green1} discovered that the gravitational and
Yang-Mills anomalies of the
ten-dimensional superstrings all cancel provided the gauge group is either
$SO(32)$ or $E_8  \times
E_8$;

2) Gross, Harvey, Martinec and Rohm \cite{Harvey} discovered the heterotic
(hybrid) string with the
above gauge groups;

3) Candelas, Horowitz, Strominger and Witten \cite{Candelas} discovered that
the  $E_8  \times
E_8$ heterotic string admits spontaneous compactifciation to four dimensions on
a six-dimensional
Calabi-Yau manifold. The resulting four-dimensional theory resembles a GUT
theory based on the
group $E_6$.  In particular, there are chiral families of quarks and leptons.

This is probably the place to admit the supreme irony of the Kaluza-Klein
unification story. It is
that the heterotic superstring \cite{Harvey}, which is currently the favorite
way to unify
gravity with the other forces, while making use of (and indeed demanding) extra
spacetime dimensions
a la Kaluza-Klein, nevertheless eschews the route of getting the Yang-Mills
gauge group from the
general coordinate group.  Instead, the Yang-Mills fields are already present
in the $D=10$
dimensional formulation!  Moreover, the favorite way subsequently to compactify
the theory to $D=4$
invokes a Calabi-Yau manifold \cite{Candelas}; an $M_6$ with no isometries at
all! Does this mean
that the idea of non-abelian gauge fields from extra dimensions is dead?
Actually, no.  The
heterotic string construction involves taking the right-moving modes on the
$2$-dimensional
worldsheet to correspond to a $10$-dimensional superstring and the left-moving
modes to correspond
to a $26$-dimensional bosonic string. The $D=10$ formulation is obtained by
compactifying the extra
$16$ dimensions on a specially chosen $T^{16}$ corresponding to an even,
self-dual lorentzian
lattice.  This leads uniquely to the anomaly-free \cite{Green1}, dimension
$496$, gauge groups $E_8
\times E_8$ or $SO(32)$.  The Cartan subalgebra $U(1)^{16}$ admits a
Kaluza-Klein interpretation,
but the remaining $480$ gauge bosons arise as solitons from the very different
Frenkel-Kac
mechanism. However, from the point of view of conformal field theory, a string
moving on such a
$k$-torus is indistinguishable from a string moving on a simply laced group
manifold $G$ with
$k=rank{} G$.  In fact, therefore, it is possible to understand {\it all} of
the $496$ gauge bosons
as Kaluza-Klein gauge bosons having arisen from compactifcation on the group
manifold
\cite{Nilsson3,Nilsson4}. (One gets Yang-Mills fields of $G$ rather than $G
\times G$ because the
construction applies only to the left-movers and not the right-movers.)  It has
to be admitted,
however, that such a traditional Kaluza-Klein interpretation, though valid, has
not lead to any new
insights into string theory.

Nevertheless, many other discoveries of the Kaluza-Klein supergravity era do
continue to influence
current thinking in string theory. These include: the connection between
holonomy, Killing spinors
and unbroken supersymmetry in four dimensions; Calabi-Yau manifolds (first
proposed as a way of
going from 10 to 6 on K3); orbifolds; fermion condensates and the cosmological
constant; the
higher dimensional interpretation of the Higgs mechanism as a distortion of the
extra dimensional
geometry; the importance of topology, index theorems and zero modes in
determining the massless
spectrum.  All this work is summarized in the Physics Report by Duff, Nilsson
and Pope
\cite{Nilsson2}, where the corresponding references may be found.

Historical reviews of string theory frequently dwell on the early days of Regge
theory, the Veneziano
model and the dual resonance model of hadrons and then jump to 1984 as though
nothing much happened
in between. In my opinion, however, the renaissance of string theory in 1984
owed more to
Kaluza-Klein supergravity than it did to the dual resonance model.

In addition to Calabi-Yau manifolds many sophisticated ways of arriving at a
consistent
four-dimensianal heterotic string theory have been studied in the last ten
years, including:
self-dual Lorentzian lattices, symmetric and asymmetric orbifolds, fermionic
formulations, etc.
These were reviewed in \cite{Duff4} and I will not repeat the story here.
Suffice it to say that
despite a good deal of technical progress, we are still no closer to resolving
what is perhaps
the most important issue in string theory, namely the vacuum degeneracy
problem.
Finding the right compactification has become synonomous with finding the right
$c=9$
superconformal field theory and there are literally billions (maybe an infinite
number?) of
candidates. For the time being therefore, the phrase {\it superstring inspired
phenomenology} can
only mean sifting through these billions of heterotic models in the hope of
finding one that is
realistic.  The trouble with this needle-in-a-haystack approach is that even if
we found a model
with good phenomenology, we would be left wondering in what sense this could be
described as a {\it
prediction} of superstrings.

The general consensus now is that this problem will never be resolved provided
we remain within the
confines of a weak coupling perturbation expansion.  I would therefore like to
finish with some
very recent developments in string theory which address this strong coupling
problem, and I am
delighted to say that they rely heavily on the Kaluza-Klein idea.

\section{Kaluza-Klein states as extreme black holes}
\la{strings}

The idea that elementary particles might behave like black holes is not a
new one \cite{Hawking,Salam2,tHooft}. Intuitively, one might expect that
a pointlike object whose mass exceeds the Planck mass, and whose Compton
wavelength is therefore less than its Schwarzschild radius, would exhibit
an event horizon. In the absence of a consistent quantum theory of gravity,
however, such notions would always remain rather vague. Superstring theory,
on the other hand, not only predicts such massive states but may provide us
with a consistent framework in which to discuss them. In this section we shall
summarize the
results of \cite{Rahmfeld} and confirm the claim \cite {Khuri} that certain
massive
excitations of four-dimensional superstrings are indeed black holes. Our
results thus complement
those of \cite{Ellis,Susskind,Russo} where it is suggested that all black holes
are single string
states. Of course, non-extreme black holes
would be unstable due to the Hawking effect. To describe stable elementary
particles, therefore, we must focus on extreme black holes whose masses
saturate a Bogomol'nyi bound.

Here we return to the humble torus and consider the four-dimensional heterotic
string
obtained by toroidal compactification. At a generic point in the moduli
space of vacuum configurations the unbroken gauge symmetry is $U(1)^{28}$
and the low energy effective field theory is described by the $N=4$
supergravity
coupled to the $22$ abelian vector multiplets of section (\ref{torus}).
We shall consider the Schwarz-Sen \cite{Schwarz,Sen} $O(6,22;Z)$ invariant
spectrum of elementary electrically charged massive $N_R=1/2,N_L=1$ states of
this
four-dimensional heterotic string, and show that the spin zero states
correspond
to extreme limits of the Kaluza-Klein black hole solutions of section
(\ref{black}) which preserve
$1/2$ of the spacetime supersymmetries. By supersymmetry, the black hole
interpretation then
applies to all members of the $N=4$ supermultiplet \cite{Gibbons2,Aichelburg},
which has
$s_{max}=1$. Here $N_L$ and $N_R$
refer to the number of left and right oscillators respectively. The $N=4$
supersymmetry algebra
possesses two central charges $Z_1$ and $Z_2$.  The $N_R=1/2$ states correspond
to that subset of
the full spectrum that belong to the $16$ complex dimensional ($s_{max}\geq1$)
representation of
the $N=4$ supersymmetry algebra, are annihilated by half of the supersymmetry
generators and
saturate the strong Bogomol'nyi bound $m=|Z_1|=|Z_2|$. As discussed in
\cite{Witten,Osborn,Schwarz,Sen}, the reasons for focussing on this N=4 theory,
aside from its
simplicity, is that one expects that the allowed spectrum of electric and
magnetic charges is not
renormalized by quantum corrections, and that the allowed mass spectrum of
particles saturating
the Bogomol'nyi bound is not renormalized either.

Following \cite{Duff1,Duff2,Font}, Schwarz and Sen have also conjectured
\cite{Schwarz,Sen} on the basis of string/fivebrane duality
\cite{Duff5,Strominger} that,
when the solitonic excitations are included, the full string spectrum is
invariant not only under the target space $O(6,22;Z)$ ($T$-duality) but also
under the strong/weak coupling $SL(2,Z)$ ($S$-duality). The
importance of $S$-duality in the context of black holes in string theory has
also been stressed in \cite{Kalara}. Schwarz and Sen have
constructed a manifestly $S$ and $T$ duality invariant mass spectrum.
$T$-duality transforms electrically charged winding states into electrically
charged Kaluza-Klein states, but $S$-duality transforms elementary
electrically charged string states into solitonic monopole and dyon states.
We shall show that these states are also described by the extreme magnetically
charged
black hole solutions discussed in \cite{Khuri}. Indeed, although the results of
this section
may be understood without resorting to string/fivebrane duality, it
nevertheless
provided the motivation. After compactification from $D=10$ dimensions to
$D=4$, the solitonic fivebrane solution of $D=10$ supergravity \cite{Lu2}
appears as a magnetic monopole \cite{Hmono,Gauntlett2} or a string
\cite{Khuri2} according as it
wraps  around $5$ or $4$ of the compactified directions \footnote{It could in
principle also appear
as a membrane by wrapping around $3$ of the compactified directions, but the
$N=4$ supergravity
theory (\ref{3}) obtained by naive dimensional reduction does not admit the
membrane
solution \cite{Khuri2}.}.
Regarding this dual string as fundamental in its own right interchanges the
roles of $T$-duality and $S$-duality. The solitonic monopole states obtained
in this way thus play the same role for the dual string as the elementary
electric winding states play for the fundamental string. The Kaluza-Klein
states are common to both. Since these solitons are extreme
black holes \cite{Khuri}, however, it follows by $S$-duality that the
elementary Kaluza-Klein states should be black holes too! By $T$-duality,
the same holds true of the elementary winding states. Rather than invoke
$S$-duality,
 however, we shall proceed directly to establish that the elementary states
described above are in
one-to-one correspondence with the extreme electric black holes \footnote{The
idea that
there might be a dual theory which interchanges Kaluza-Klein states and
Kaluza-Klein monopoles
was previously discussed in the context of $N=8$ supergravity by Gibbons and
Perry
\cite{Gibbons3}.}. Now this leaves open the possibility that they have the same
masses and quantum
numbers but different interactions. Although we regard this possibilty as
unlikely given the
restrictions of $N=4$ supersymmetry, the indirect argument may be more
compelling in this respect
(even though it suffers from the drawback that $S$-duality has not yet been
rigorously
established).  Of course, elementary states are supposed to be singular and
solitonic states
non-singular. How then can we interchange their roles? As we saw in section
(\ref{black}), the way
the theory accommodates this requirement is that when expressed in terms of the
fundamental metric
$e^{\sqrt 3\phi}g_{\mu\nu}$ that couples to the worldline of the superparticle
the elementary
solutions are singular and the solitonic solutions are non-singular, but when
expressed in terms of
the dual metric $e^{-\sqrt 3\phi}g_{\mu\nu}$, it is the other way around
\cite{Lu1,Khuri}.

We now turn to the electric and magnetic charge spectrum. Schwarz
and Sen \cite{Schwarz,Sen} present an $O(6,22;Z)$ and $SL(2,Z)$ invariant
expression
for the mass of particles saturating the strong Bogomol'nyi bound
$m=|Z_1|=|Z_2|$:
\be
m^2=\frac{1}{16}(\alpha^a~~\beta^a){\cal M}^{0}(M^{0}+L)_{ab}
    \left(\begin{array}{c}
\alpha^b\\
\beta^b\end{array}\right)
\la{5}
\ee
where a superscript $0$ denotes the constant asymptotic values of the
fields. Here $\alpha^a$ and $\beta^a$ ($a=1,...,28$) each belong to an even
self-dual Lorentzian lattice $\Lambda$ with metric given by $L$ and are
related to the electric and magnetic charge vectors $(Q^a,P^a)$ by
$(Q^a,P^a)=\left(M_{ab}{}^0(\alpha^b + \lambda_{1}{}^0\beta^b)/\lambda_2{}^0,
L_{ab}\beta^b\right)$.  The fundamental charge $Q$ is normalized so that
$Q^2=e^2/2 \pi$ and we
have set $\kappa^2=16\pi$.  As discussed in \cite{Schwarz,Sen} only a subset of
the conjectured
spectrum corresponds to elementary string states.  First of all these states
will be only
electrically charged, i.e. $\b =0$, but there will be restrictions on $\a$ too.
After performing
an $O(6,22)$ rotation $\W$ of the background $M^0$ transforming it into the
28-dimensional identity matrix $I$ and an accompanied change of basis to
$\hat { \a}=L \W L { \a}$ the mass formula (\ref{5}) can be rewritten
\cite{Schwarz,Sen} as
\be
m^2=\frac{1}{16 \lambda_2{}^{0}}
 \hat {\alpha}^a (I+L)_{ab} \hat {\alpha}^b=
 \frac{1}{8\lambda_2{}^{0}}
 \left(\hat {\sf \alpha}_R \right)^2
\la{7}
\ee
with $\hat {\sf \alpha}_R = \frac{1}{2} (I+L) \hat {\sf \alpha}$ and $\hat
{\sf \alpha}_L = \frac{1}{2} (I-L) \hat {\sf \alpha}$. In the string
language $\hat{\sf \a}_{R(L)}$ are the right(left)-moving internal momenta.
The mass of a generic string state in the Neveu-Schwarz sector (which is
degenerate with the Ramond sector) is given by
\be
m^2=\frac{1}{8\lambda_2{}^{0}} \left\{
 \left(\hat {\sf \alpha}_R \right)^2 +2 N_R -1 \right\}=
 \frac{1}{8 \lambda_2{}^{0}} \left\{
 \left(\hat {\sf \alpha}_L \right)^2 +2 N_L -2 \right\}.
\label{8}
\ee
A comparison of (\ref{7}) and
(\ref{8}) shows that the string states
satisfying the Bogomol'nyi bound all have $N_R=1/2$. One then finds
\be
 N_L-1=\frac{1}{2} \left(\left(\hat {\sf \alpha}_L \right)^2
 -\left(\hat {\sf \alpha}_R \right)^2\right)
 =\frac{1}{2} {\sf \alpha}^{T} L {\sf \alpha},
\label{9}
\ee
leading to ${\sf \alpha}^{T} L {\sf \alpha}\geq -2$.

We shall now show that
the extreme Kaluza-Klein black holes are string states with ${\sf
\alpha}^{T} L {\sf \alpha}$ null ($N_L=1$). To identify them as states
in the spectrum we have to find the corresponding charge vector ${\sf \a}$
and to verify that the masses calculated by the formulas (\ref{bound}) and
(\ref{5}) are identical. The action (\ref{3}) can be consistently truncated
by keeping the metric $g_{\mu\nu}$, just one field strength ($F^1$ say), and
one scalar field $\phi$ via the ansatz $\Phi=\phi/\sqrt{3}$ and
$M_{11}=e^{2\phi/\sqrt{3}}=M_{77}^{-1}$. All other diagonal components of
$M$ are set equal to unity and all non-diagonal components to zero. Now
(\ref{3}) reduces to (\ref{action2}).  (This yields the electric and magnetic
Kaluza-Klein
(or "$F$") monopoles. This is not quite the truncation chosen in \cite{Khuri},
where
just $F^7$ was retained and $M_{11}=e^{-2\phi/\sqrt{3}}=M_{77}^{-1}$.  This
yields the electric
and magnetic winding (or "$H$") monopoles.  However, the two are related by
$T$-duality).  We shall
restrict ourselves to the purely electrically charged solution with charge
$Q=1$, since this one
is expected to correspond to an elementary string excitation. The charge vector
${\sf \a}$ for
this solution is obviously given by $\alpha^a=\delta^{a,1}$ with ${\sf
\alpha}^{T} L {\sf
\alpha}=0$.  Applying (\ref{5}) for the mass of the state we find
$m^2=1/16=Q^2/16$, which
coincides with (\ref{bound}) in the extreme limit.  This agreement confirms the
claim that this
extreme Kaluza-Klein black hole is a state in the Schwarz-Sen spectrum and
preserves
$2$ supersymmetries. From this solution we can generate the whole set of
supersymmetric black hole solutions with ${\sf \alpha}^{T} L {\sf \alpha}=0$
in the following way: first we perform an $O(6,22;R)$ transformation to obtain
 a vector proportional to the desired lattice vector. This clearly leaves the
mass invariant, but
the new charge vector ${\sf \a}'$ will in general not be located on the
lattice. To find a state
in the allowed charge spectrum we have to rescale ${\sf \a}'$ by a constant $k$
so that ${\sf
\a}''=k{\sf \a}'$ is a lattice vector. Clearly the masses calculated by
(\ref{bound}) and
(\ref{5}) still agree (this is obvious by reversing the steps of rotation
and rescaling), leading to the conclusion that all states obtained in this
way preserve 1/2 of the supersymmetries. Therefore all states in the
spectrum belonging to $s_{max}=1$ supermultiplets for which $N_R=1/2,N_L=1$ are
extreme
Kaluza-Klein black holes.

It should also be clear that the purely magnetic extreme black hole
solutions \cite{Khuri} obtained from the above by the replacements $
\phi\rightarrow -\phi, F\rightarrow e^{-\sqrt 3 \phi}{}^*{}F$ will also belong
to the
Schwarz-Sen spectrum of solitonic states.  Starting from either the purely
electric or purely magnetic solutions, dyonic states in the spectrum which
involve non-vanishing axion field $\Psi$ can then be obtained by $SL(2,Z)$
transformations. Specifically , a black hole with charge vector $({\sf
\alpha}, 0)$ will be mapped into ones with charges $(a{\sf \alpha}, c{\sf
\alpha})$ with the integers $a$ and $c$ relatively prime \cite{Schwarz,Sen}.

We have limited ourselves to $N_R=1/2,N_L=1$ supermultiplets with $s_{min}=0$.
Having established
that the $s=0$ member of the multiplet is an extreme black hole, one may then
use the fermionic
zero modes to perform supersymmetry transformations to generate the whole
supermultiplet of black
holes \cite{Gibbons2,Aichelburg}. Of course there are $N_R=1/2$ multiplets with
$s_{min}>0$ coming
from oscillators with higher spin and our arguments have nothing to say about
whether these are
also extreme black holes.

As discussed in {\cite{Rahmfeld} the $N_R=1/2,N_L>1$ states (and their dual
counterparts) are also
extreme electric (magnetic) black holes.  However, whereas the Kaluza-Klein
black holes have a
scalar-Maxwell coupling $e^{-a\phi}F^{\mu\nu}F_{\mu\nu}$ with $a=\sqrt 3$, the
$N_L>1$ states have
$a=1$ and correspond to the supersymmetric dilaton black hole \cite{Gibbons}.
It is well-known
that the non-supersymmetric $a=1$ case provides a solution of the heterotic
string
\cite{Garfinkle,Giddings} but it was only recently recognized that the
supersymmetric
$a=1$ case and also the $a=\sqrt 3$ case are also solutions
\cite{Khuri,Rahmfeld}. The electric
solutions are in fact {\it exact} \cite{Tseytlin} with no $\alpha'$ corrections
for both $a=\sqrt
3$ and $a=1$. There are also $a=0$ Reissner-Nordstrom black holes
\cite{Rahmfeld}, but these do not
belong to the $N_- =1/2$ sector of string states. None of the spinning
$N_R=1/2$ states is
described by extreme {\it rotating} black hole metrics because they obey the
same Bogomol'nyi
bound as the $s_{min}=0$ states, whereas the mass formula for an extreme {\it
rotating} black hole
depends on the angular momentum $J$. Rather it is the fermion fields which
carry the spin. (For
the $a=0$ black hole, they yield a gyromagnetic ratio $g=2$ \cite{Aichelburg};
the $a=\sqrt{3}$
and $a=1$ superpartner $g$-factors are unknown to us.)  It may be that there
are states in the
string spectrum described by the extreme rotating black hole metrics but if so
they will belong to
the $N_R\neq 1/2$ sector\footnote{The gyromagnetic and gyroelectric ratios of
the states in the
heterotic string spectrum would then have to agree with those of charged
rotating black hole
solutions of the heterotic string. This is indeed the case: the $N_L=1$ states
\cite{Hosoya} and
the extreme rotating $a=\sqrt{3}$ black holes \cite{Wiltshire} both have $g=1$
whereas the $N_L>1$
states \cite{Russo} and the extreme rotating $a=1$ \cite{Sen2} (and $a=0$
\cite{Schild}) black holes
both have $g=2$. In fact, it was the observation that the Regge formula $J\sim
m^2$ also describes
the mass/angular momentum relation of an extreme rotating black hole which
first led Salam
\cite{Salam2} to imagine that elementary particles might behave like black
holes!}.  Since,
whether rotating or not, the black hole solutions are still independent of the
azimuthal angle and independent of time, the supergravity theory is effectively
{\it
two-dimensional} and therefore possibly integrable. This suggests that the
spectrum should be
invariant under the larger duality $O(8,24;Z)$ \cite{Duff1,Duff2}, which
combines $S$ and $T$. The
corresponding Kac-Moody extension would then play the role of the spectrum
generating symmetry
\cite{Geroch}.

\section{Acknowledgments}
I should like to thank my former extra-dimensional collaborators Bengt Nilsson
and Chris Pope.  The
results described in section (\ref{strings}) are based on work carried out with
Ramzi Khuri, Rubin
Minasian and Joachim Rahmfeld. I am grateful to the Director and Staff of the
Isaac Newton
Institute, and to the organizers of the {\it Topological Defects} programme,
for their hospitality.
\newpage

\end{document}